\documentclass{jpsj-suppl}
\usepackage{units}
\usepackage{subfigure}
\usepackage{chngpage}

\title{Electron Neutrino Charged-Current Quasielastic Scattering in the {MINERvA} Experiment}

\author{Jeremy Wolcott, for the {MINERvA} collaboration}
\inst{Department of Physics and Astronomy, University of Rochester, Rochester, NY 14627 USA \\
Department of Physics and Astronomy, Tufts University, Medford, MA 02155 USA}

\newcommand{\mnv}{MINERvA}

\email{jwolcott@fnal.gov}

\recdate{December 21, 2015}

\abst{The electron-neutrino charged-current quasielastic (CCQE) cross section on nuclei is an important input parameter for electron neutrino appearance oscillation experiments.  Current experiments typically begin with the muon neutrino cross section and apply theoretical corrections to obtain a prediction for the electron neutrino cross section.   However, at present no experimental verification of the estimates for this channel at an energy scale appropriate to such experiments exists.  We present the cross sections for a CCQE-like process determined using the \mnv{} detector, which are the first measurements of any exclusive reaction in few-GeV electron neutrino interactions.  The result is given as differential cross-sections vs. the electron energy, electron angle, and square of the four-momentum transferred to the nucleus, $Q^2$.  We also compute the ratio to a muon neutrino cross-section in $Q^{2}$ from \mnv{}.  We find satisfactory agreement between these measurements and the predictions of the GENIE generator.  We furthermore report on a photon-like background unpredicted by the generator which we interpret as neutral-coherent diffractive scattering from hydrogen.}

\kword{electron neutrino, cross section, quasielastic, diffractive}

\begin{document}
\maketitle

	\section{Introduction}
		Current terrestrial neutrino oscillation experiments searching for fundamental information in the neutrino sector, such as the neutrino mass ordering and whether CP violation occurs for leptons, usually employ experimental designs which rely on the partial oscillation of a beam of muon neutrinos into electron neutrinos.\cite{T2K NIM,NOvA TDR}  These experiments build large detectors of heavy materials to maximize the rate of neutrino interactions, and then examine the energy distribution of the neutrinos that do interact with the detector, comparing the observed spectrum with predictions based on hypotheses of no oscillation or oscillation with given parameters.
		
		Correctly predicting the observed energy spectrum is necessary for the inference of the magnitude and shape of neutrino oscillations.  To do this, one must be able to accurately model the rates and outgoing particle kinematics.  Thus, one needs precise $\nu_{e}$ cross sections on the detector materials in use.  And yet, because of the difficulties associated with producing few-GeV electron neutrino beams, there are very few such cross section measurements\cite{Gargamelle nue,T2K nue}.  Furthermore, the small statistics and inclusive nature of these measurements make it challenging to use them for tuning or evaluating the models used in simulation programs.  Instead, modern simulations begin from the wealth of high-precision cross-section data available for muon neutrinos and apply corrections such as those discussed in ref. \cite{DayMcF} to obtain a prediction for $\nu_{e}$.
		
		This analysis presents a higher-statistics cross section for a quasielastic-like electron neutrino process, which is among the dominant reaction mechanisms at most energies of interest to oscillation experiments.  We use the \mnv{} detector, which consists of a central sampling scintillator region, built from strips of fluoror-doped scintillator glued into sheets, then stacked transverse to the beam axis; both barrel-style and downstream longitudinal electromagnetic and hadronic sampling calorimeters; and a collection of upstream passive targets of lead, iron, graphite, water, and liquid helium.  The detector design and performance are discussed in full detail elsewhere.\cite{MINERvA NIM}  \mnv{} is situated in the NuMI $\nu_{\mu}$ beam, where it was exposed to a flux of $\sim 99$\% $\nu_{\mu}$ and $\sim 1$\% $\nu_{e}$ mostly between \unit[2-5]{GeV} for this data set.	We also compare the result for $\nu_{e}$ to a similar, previous \mnv{} result for $\nu_{\mu}$ to evaluate the assumption of the model that the only relevant difference between $\nu_{\mu}$ and $\nu_{e}$ charged-current scattering is due to the mass of the final-state charged lepton.
	
	\section{Signal definition}

		In traditional charged-current quasielastic neutrino scattering, CCQE, the neutrino is converted to a charged lepton via exchange of a W boson with a nucleon, resulting in the following reaction: $\nu_{l} n \rightarrow l^{-} p$.  (Antineutrino scattering reverses the lepton number and isospin: $\bar{\nu}_{l} p \rightarrow l^{+} n$.)  Because the \mnv{} detector is not magnetized, in this analysis we cannot differentiate between electrons and positrons on an event-by-event basis.  Moreover, hadrons exiting the nucleus after the interaction can re-interact and change identity or eject other hadrons\cite{GiBUU FSI}; furthermore, interactions between nucleons within the initial state may cause multiple nucleons to be ejected by a single interaction or deform the observed kinematics\cite{Martini corr,Nieves corr}.  Therefore, we choose a signal definition for this analysis that is more closely related to what is observable in the experiment: we search for events with either an electron or positron, no other leptons or photons, any number of nucleons, and no other hadrons, irrespective of the ``true'' original process.  We call this type of event ``CCQE-like.''  We also demand that events originate from a 5.57-ton volume fiducial volume in the central scintillator region of \mnv{}.
		
	\section{Event selection and working sample}
		\label{sec:selection}
		
\begin{figure}[htb]
			\centering
			
			\subfigure[]{%
				\includegraphics[width=0.45\textwidth]{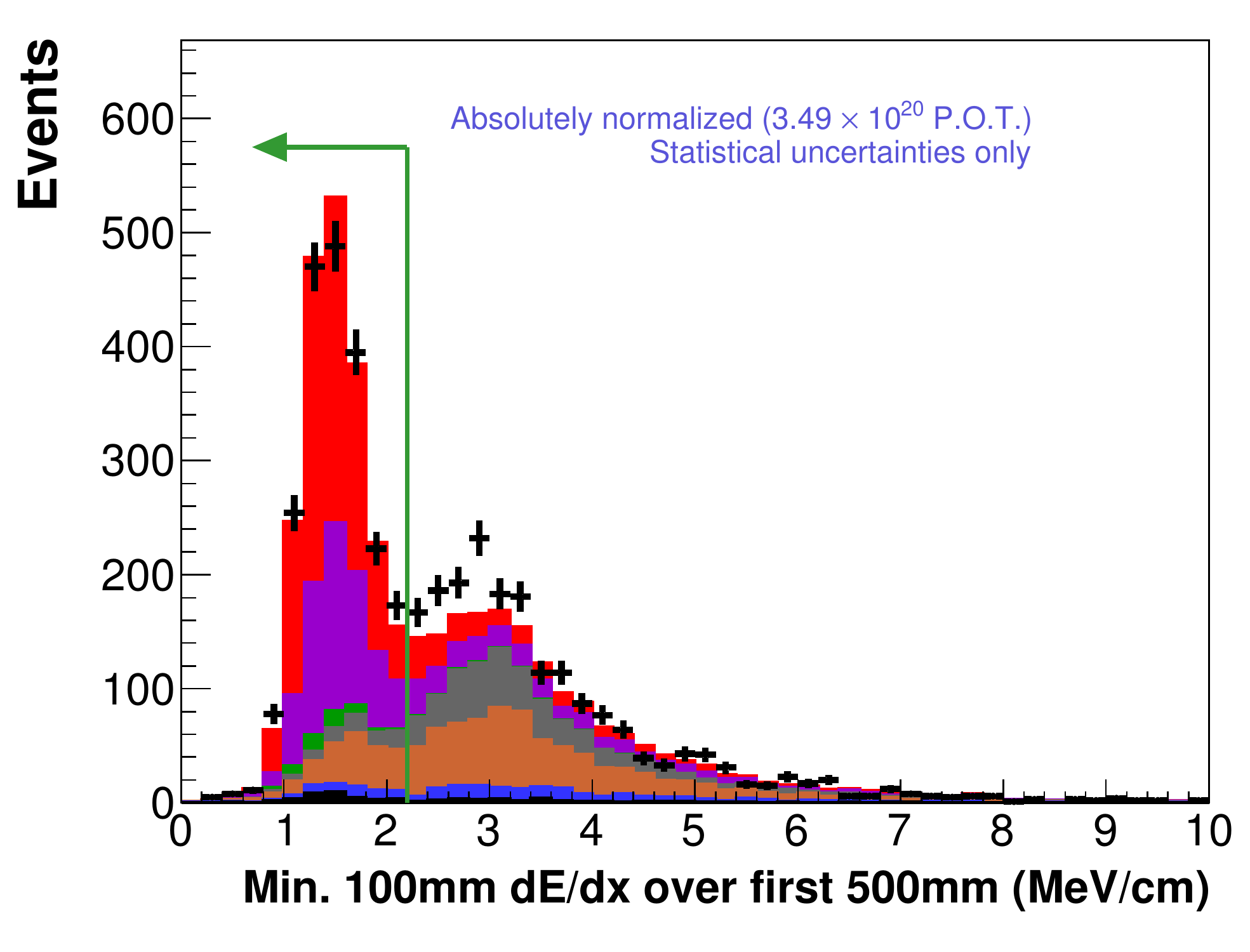}
				\label{fig:dEdx}
			}
			\subfigure[]{%
				\includegraphics[width=0.45\textwidth]{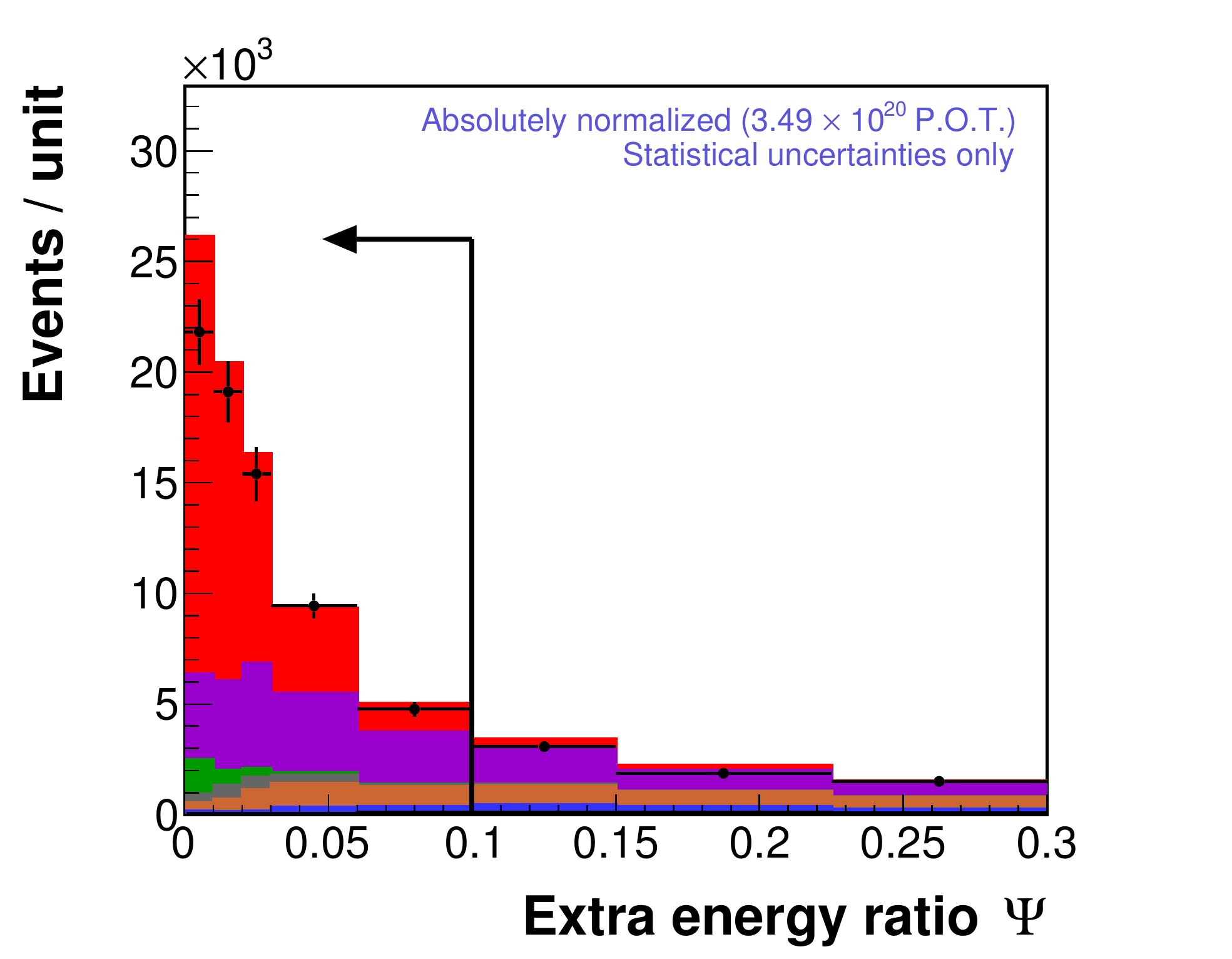}
				\label{fig:psi}
			}
			\subfigure[]{%
				\centering
				\includegraphics[width=0.5\textwidth]{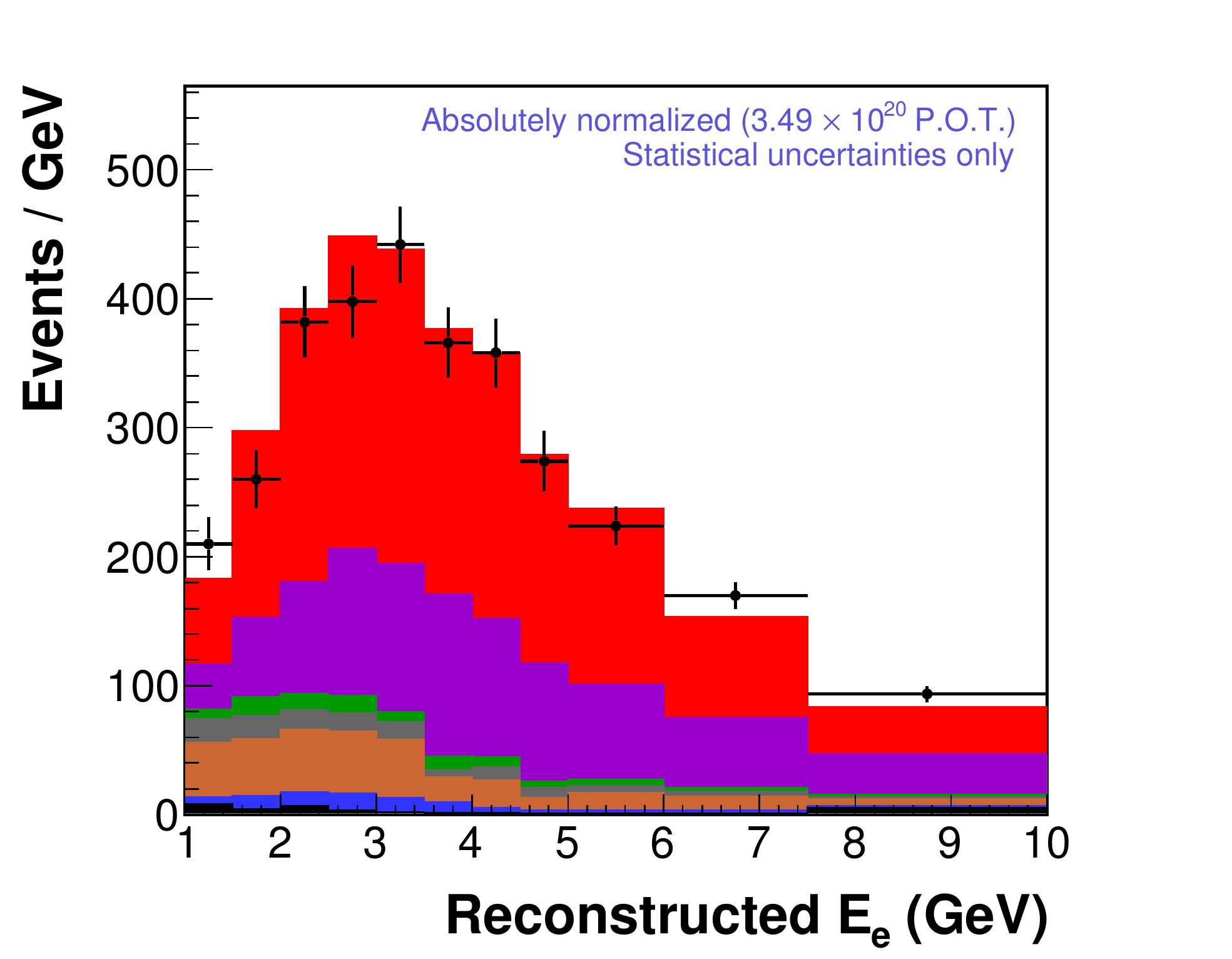}
				\label{fig:selected sample}
			}
			\subfigure{%
				\centering
				\includegraphics[width=0.75\textwidth]{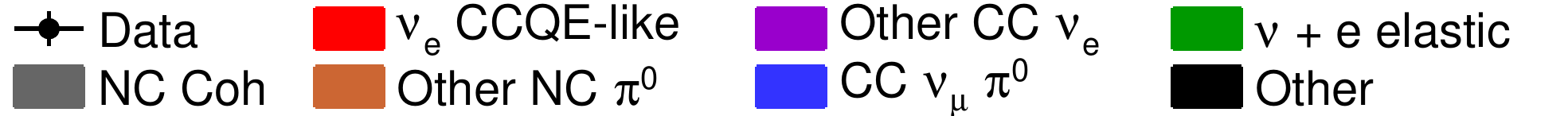}
			}
			\caption{Top left: cut on minimum front $dE/dx$.  Top right: example cut on $\Psi$ (defined in the text) at the most probable event visible energy, $E_{\mathrm{vis}} = \unit[1.25]{GeV}$.  Bottom: event sample (vs. reconstructed electron energy) after all selection cuts.}
			\label{fig:selection}
		\end{figure}

		Candidate events are selected from the data based on four major criteria.  First, a candidate must contain a reconstructed electromagnetic shower primarily contained within a cone of opening angle $7.5^{\circ}$, originating in the fiducial volume, which is identified as a shower by a multivariate particle identification algorithm.  The latter combines details of the energy deposition pattern both longitudinally (mean $dE/dx$, fraction of energy at downstream end of cone) and transverse to the axis of the cone (mean shower width) using a $k$-nearest-neighbors (kNN) algorithm.  Secondly, we separate electrons and positrons from photons by rejecting events in which the energy deposition rate ($dE/dx$) at the upstream end of the shower is consistent with two particles rather than one (since photons typically interact in \mnv{} by producing an electron-positron pair), shown in fig. \ref{fig:dEdx}.  At this point, showers surviving the cuts become electron candidates.  Thirdly, we remove events with candidate muon decay electrons identified by their separation in time from the main event; these Michel electrons typically occur in inelastic interactions with final-state pions ($\pi^{\pm} \rightarrow \mu^{\pm} \rightarrow e^{\pm}$).  Our final criterion is an attempt to select CCQE-like interactions using a variable we call ``extra energy fraction,'' $\Psi$.  Denoting an event's visible energy not inside the electron candidate, a reconstructed track beginning at the cone vertex, or a sphere of radius \unit[30]{cm} centered around the cone vertex ``extra energy,'' the extra energy fraction is defined as:
		\begin{equation} \Psi = \frac{E_{\mathrm{extra}}}{E_{\mathrm{electron}}} \end{equation}
		Our cut on $\Psi$ is a function of the total visible energy of the event.  The cut at the most probable total visible energy, $E_{\mathrm{vis}} = \unit[1.25]{GeV}$, is illustrated in fig. \ref{fig:psi}.  Finally, we retain only events with reconstructed electron energy $E_{e} \geq \unit[0.5]{GeV}$ and reconstructed neutrino energy $E_{\nu}^{QE} \leq \unit[10]{GeV}$.  Here the lower bound excludes a region where the expected flux of electron-flavor neutrinos is small and the backgrounds are large, and the upper bound restricts the sample to events where the uncertainties on flux prediction are tolerable.  The distribution of events selected by this sequence is shown in fig. \ref{fig:selected sample}.
		
		\section{Backgrounds}
		
			As fig. \ref{fig:selected sample} shows, even after the final selection, a significant fraction of the sample is predicted to be from background processes.  To validate and constrain the background predictions from the generator, we begin by using an \textit{in situ} \mnv{} measurement based on elastic scattering of neutrinos from atomic electrons\cite{MINERvA nu+e} and a recent \mnv{} measurement of charged-current coherent pion production\cite{MINERvA coherent} to constrain the $\nu-e$ and NC coherent backgrounds.
			
			We then attempt to constrain the remaining components of the background model by examining sidebands in two of the variables already mentioned.  The first of these is composed of events that contain Michel electron candidates, which results in a nearly pure sideband of inelastic $\nu_{e}$ events.  The second sideband is in the extra energy fraction $\Psi$; a sample of events at larger $\Psi$ constitutes a sideband rich in both the $\nu_{e}$ inelastic background and backgrounds where photon(s) from a $\pi^{0}$ decay comprise the electromagnetic shower.  We use these sidebands together to fit the normalizations of the three major backgrounds: $\nu_{e}$ inelastic events, neutral-current incoherent $\pi^{0}$ events, and charged-current incoherent $\pi^{0}$ events.  The normalizations of the $\nu_{e}$ background and the sum of the $\pi^{0}$ backgrounds are each fitted using distributions in both reconstructed candidate electron angle and energy, across the two sidebands, to obtain scale factors that represent the best estimate of the normalizations in the data as compared to the prediction from GENIE.  We obtain scale factors of $0.89 \pm 0.08$ and $1.06 \pm 0.12$, respectively. 
			
			The disagreement evident in fig. \ref{fig:dEdx} between minimum front $dE/dx$ values of 2.2 and \unit[3.4]{MeV/cm} suggests that there is a process producing a photon-like final state present in the data but not modeled by the generator.  We explored various scenarios for the identity of this data excess.  We rule out mismodeling of the electromagnetic cascade process resulting in a photon-like signal because the analogous Michel electron sideband in the excess region to that considered above is in very good agreement with the prediction, whereas if the shower model were responsible for the excess, it would not be.  Furthermore, the shapes of energy profile distributions resulting from subtracting the predicted backgrounds in this region from the data are indicative of the properties of the underlying process.  By comparing these to the corresponding shapes using samples of simulated particles with the same kinematic spectra as the excess, we are able to conclude that the excess is due to the decay of neutral pions into pairs of photons.  We find the excess to be characterized by events with low $\Psi$, i.e., events with very little ``extra energy'' in them, suggesting that nearly all of the energy is contained in the photon showers arising from the $\pi^{0}$ decay.  This, in turn, implies that they arise from neutral-current interactions and from a process trasferring very little energy or momentum to the nucleus.
			
\begin{figure}
			\centering
			
			\begin{adjustwidth}{-0.5in}{-0.5in}
				\subfigure[]{%
					\includegraphics[width=0.41\textwidth]{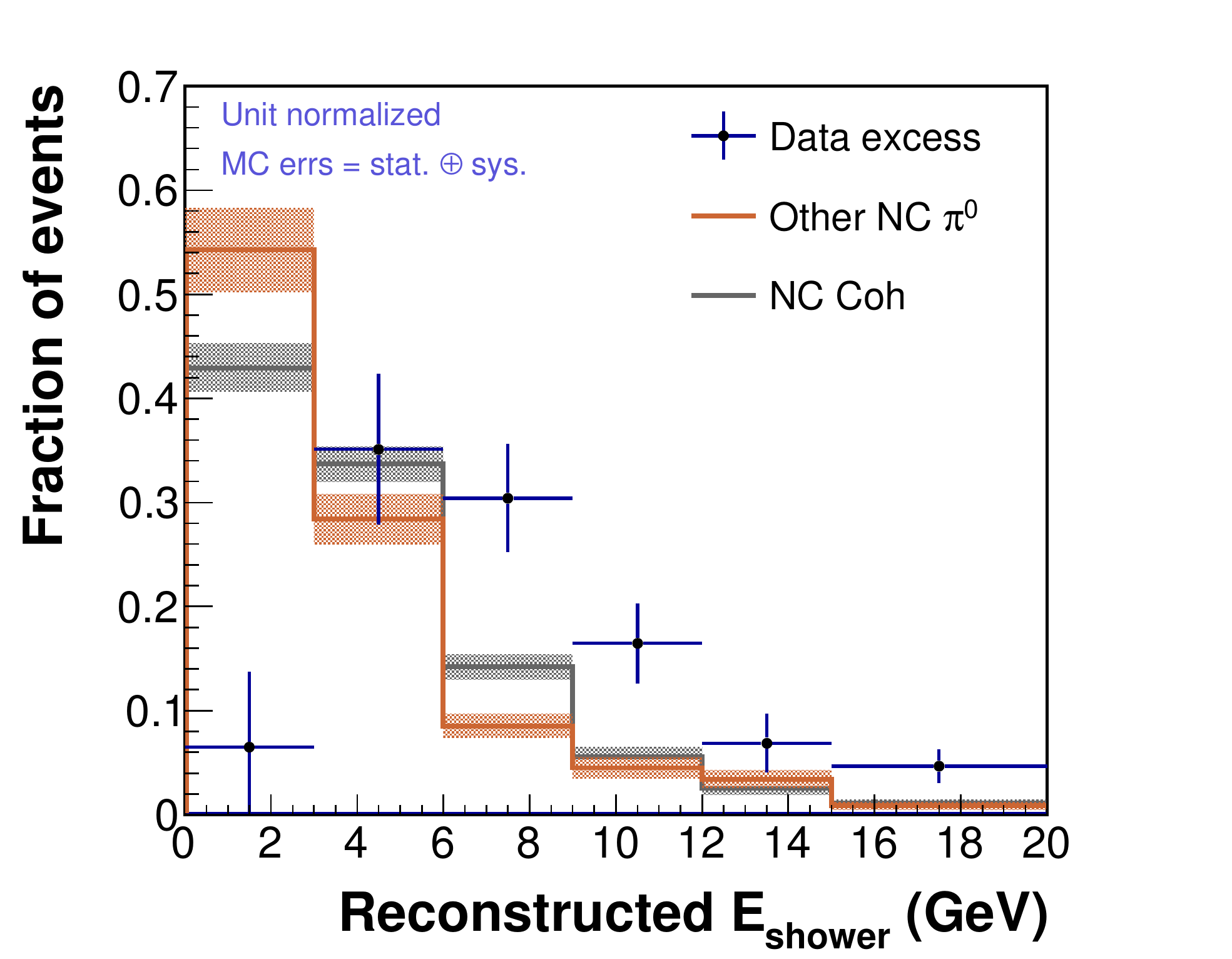}
					\label{fig:excess Eshower}
				}%
				\subfigure[]{%
					\includegraphics[width=0.41\textwidth]{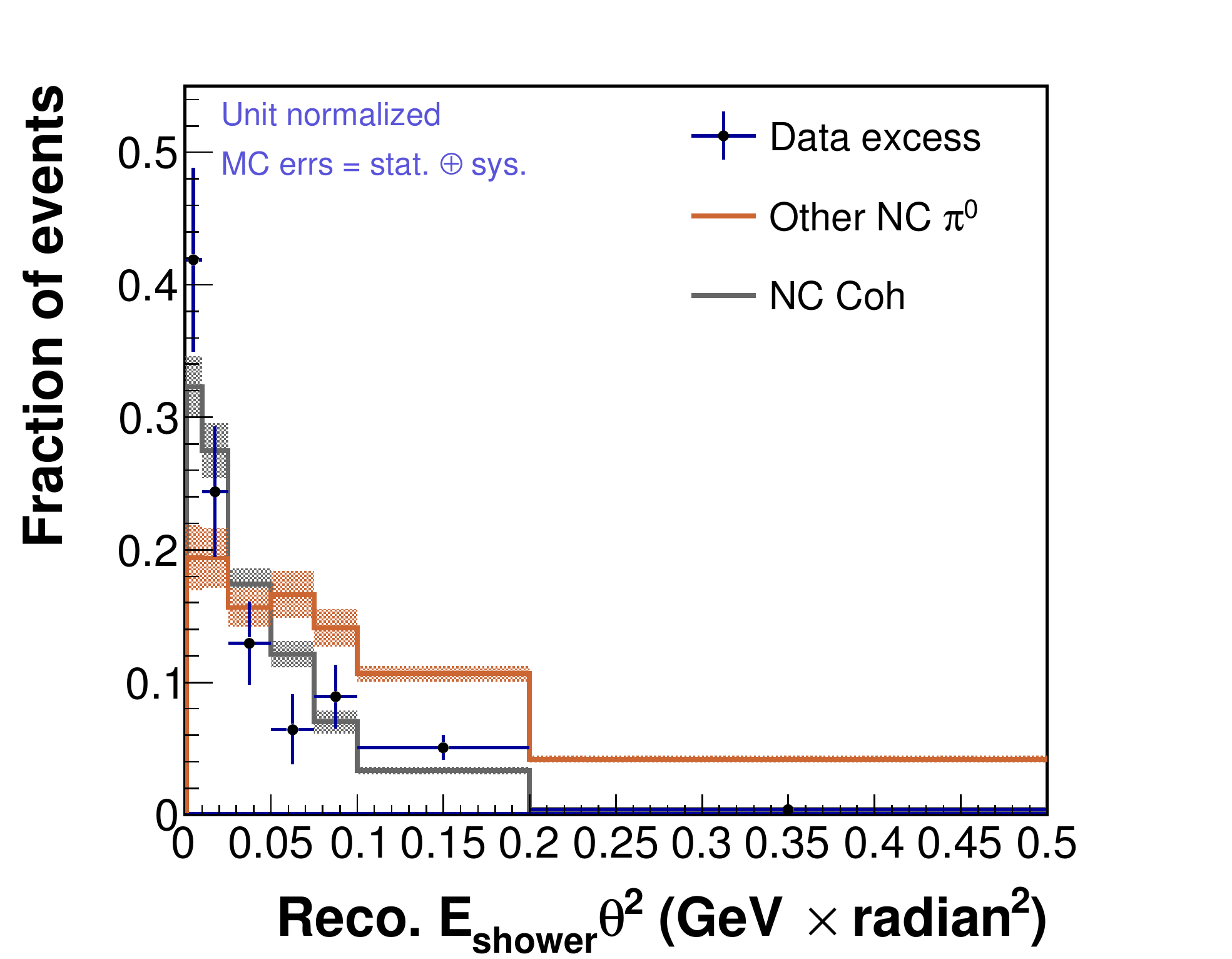}
					\label{fig:excess Etheta2}
				}%
				\subfigure[]{%
					\includegraphics[width=0.41\textwidth]{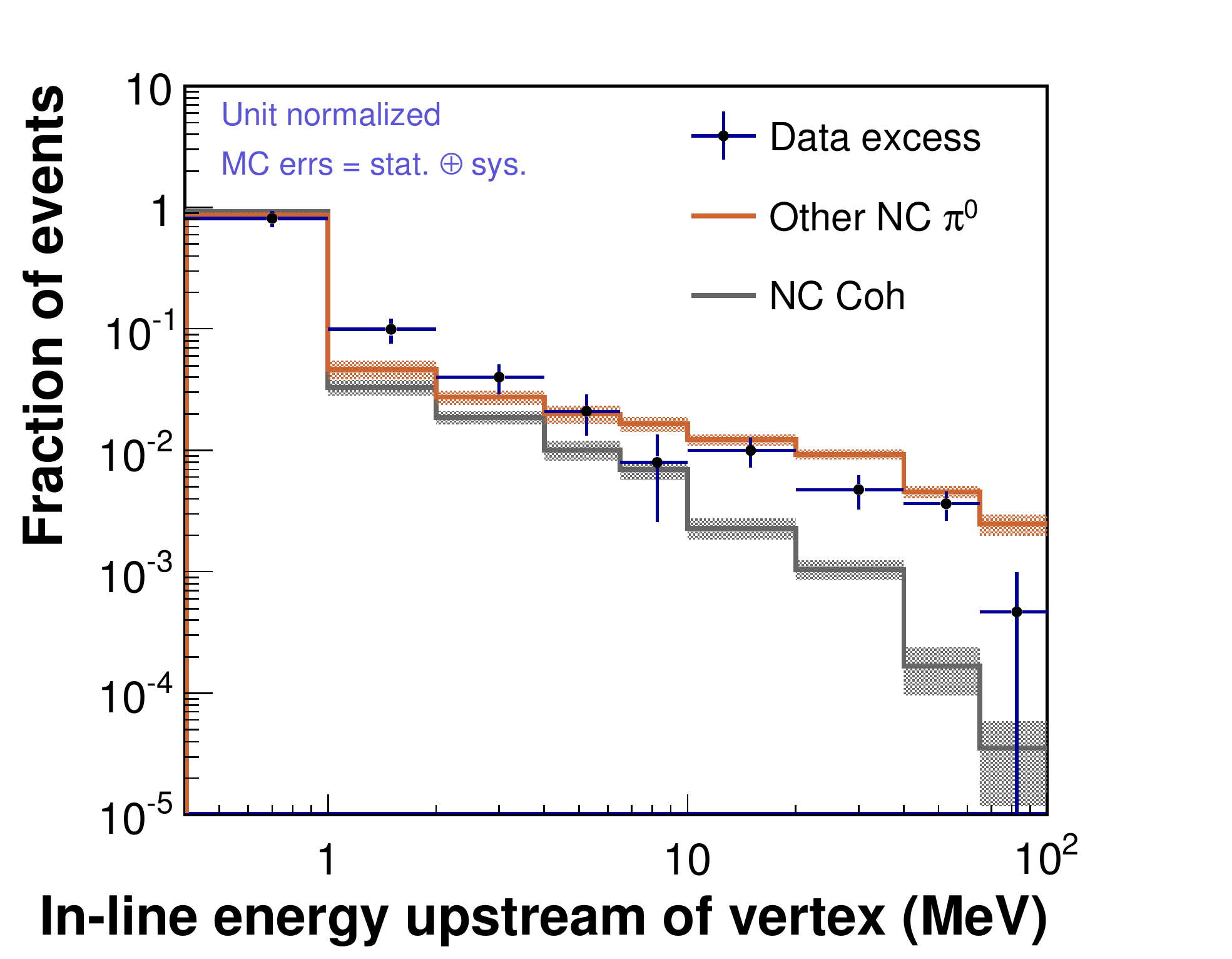}
					\label{fig:excess UIE}
				}
			\end{adjustwidth}
			\caption{Left: shape of the data excess in reconstructed shower energy compared to neutral-current predictions.  Middle: shape of the data excess in $E_{\mathrm{shower}}\theta_{\mathrm{shower}}^2$ compared to neutral-current models within the prediction.  Right: shape of the data excess in upstream inline energy (described in the text) compared to neutral-current predictions. }
		\end{figure}

			We find that the excess occupies a much more energetic region of phase space than any of our neutral-current models predict, as shown in fig. \ref{fig:excess Eshower}, which obliges us to consider the energy dependence of any other characteristics we wish to compare to our simulation.  The variable $E_{\pi}(1-\cos\theta_{\pi}) \approx E_{\pi}\theta_{\pi}^{2}$ is known to be an approximately energy-dependent measure of the ``forwardness'' of a neutrino interaction producing a pion.  It is frequently used to help identify coherent pion production from nuclei, where the wavelength corresponding to the momentum transfer (and from that, $E_{\pi}(1-\cos\theta_{\pi})$) must be of order the nuclear size and therefore has an upper bound\cite{Lackner coh, MB coh}.  In this variable, our data excess compares favorably to the neutral-current coherent process predicted by GENIE, as illustrated in fig. \ref{fig:excess Etheta2}.  However, if we look for evidence of nuclear activity upstream of the shower by summing the energy in a cone originating at the reconstructed vertex but antiparallel to the electron candidate cone described in sec. \ref{sec:selection}, we find (see fig. \ref{fig:excess UIE}) that the excess is more compatible with an incoherent process than the coherent one.  This, combined with the fact that (as noted above) MINERvA's charged-current measurement strongly disfavors the scale factor of roughly two required to make GENIE's NC coherent responsible for the excess, requires us to look elsewhere for the origin.  In fact, these features---strongly forward neutral-current scattering producing a neutral pion, some evidence of nuclear activity, and nothing else---are most consistent with a different interaction analogous to coherent scattering but occurring on hydrogen instead of heavy nuclei.  In this type of interaction, the Z-boson exchange still fluctuates into the vector meson $\pi^{0}$, as in coherent scattering.  However, because the single proton making up the hydrogen nucleus is much less massive than a bound nucleus (e.g., carbon), the four-momentum transfer often endows the proton with enough recoil kinetic energy to be observed in MINERvA.  Diffractive production, as it is sometimes known, is not simulated by GENIE's default tune (though it has an unvetted, pre-production implementation of a model for it by D. Rein\cite{Rein diffr}), which explains why we observe it as an excess relative to the prediction.  (Though neutral-current excitation of a $\Delta^{+}$ from a proton within a nucleus produces the same final state after the decay $\Delta^{+} \rightarrow p^{+}\pi^{0}$, the latter process is characterized by a strong peak around \unit[1.2]{GeV} in the invariant mass spectrum of the events.  We computed the invariant mass distribution for the excess, using the upstream inline energy distribution to form a rough estimate for the proton kinetic energy, and found a broad $W$ spectrum peaking at about \unit[3.5]{GeV} with FWHM of about \unit[3]{GeV}.  Therefore we rule out resonant production.)  We identify the excess with this diffractive process.
	
\begin{figure}
				\centering
				\includegraphics[width=0.6\textwidth]{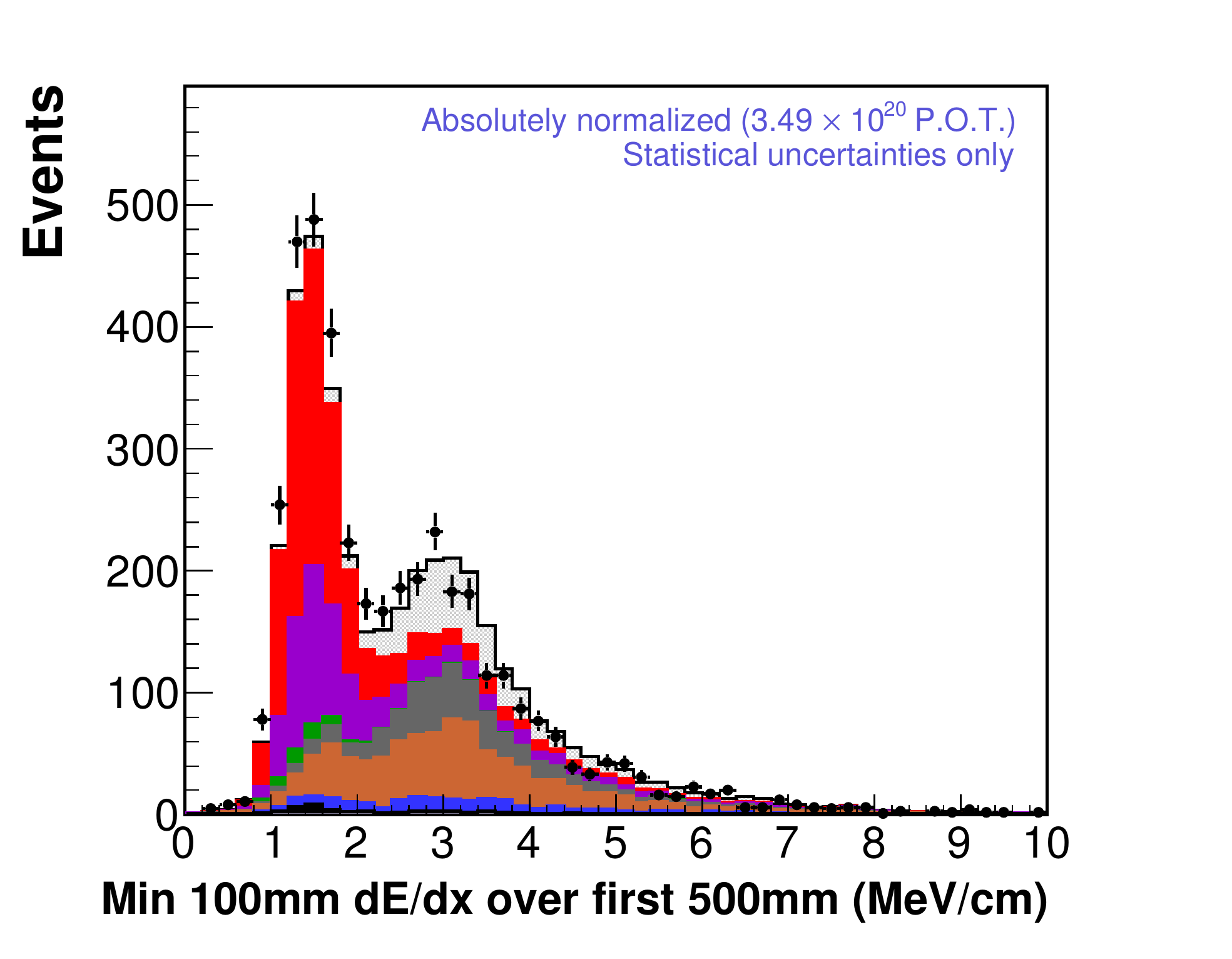}
				\caption{The minimum front $dE/dx$ distribution after the background constraint procedure (compare fig. \ref{fig:dEdx}) and including the \textit{ad hoc} model for diffractive production described in the text (top gray shaded distribution).  (The other colors are the same as in fig. \ref{fig:selection}.)}
				\label{fig:dEdx with excess model}
			\end{figure}

			The excess is incompatible with the processes already predicted by the simulation.  Therefore, in order to control for any effect it has on the background prediction in the electron neutrino signal region, we construct a prediction based on the conclusions above.  We construct an \textit{ad hoc} sample of events consisting of neutral pions with a joint reconstructed energy and angle distribution fitted to what was measured in the excess.  (The proton recoil energy corresponding to the upstream inline energy in fig. \ref{fig:excess UIE} is minimal.  Because it is in general separated from the electromagnetic shower---whose photons have traveled, unseen, at least a radiation length ($\sim \unit[40]{cm}$) on average from the nucleus where the $\pi^{0}$ was created---it does not influence the reconstruction of any of the quantities used in the results below.  We therefore neglect it.)  We add this sample to the prediction for our backgrounds, and, because this could in principle affect the scale factors mentioned above, we perform the scale factor constraint procedure again.  As there is very little contribution from the excess in the electron signal region, as shown in fig. \ref{fig:dEdx with excess model}, the scale factors change negligibly.
			
			 Subsequent to all these constraints, we scale the backgrounds in the signal region and subtract them from the data.  We then compare the simulated prediction of the signal process to the background-subtracted data.

	\section{Cross section results}
		We calculate three differential cross sections in electron angle, electron energy, and four-momentum transferred from neutrino to nucleus $Q^{2}$.  For $Q^{2}$, we employ the commonly-used CCQE approximations (assuming a stationary target nucleon) which allow us to compute the neutrino kinematics from just the lepton variables (and the masses $m_{p}$, $m_{n}$, $m_{e}$ as well as an effective binding energy for the ejected nucleon $E_{b}$):
		\begin{equation}
			E_{\nu}^{QE} = \frac{m_{n}^{2} - (m_{p} - E_{b})^{2} - m_{e}^{2} + 2(m_{p}-E_{b}E_{e})}{2(m_{p} - E_{b} - E_{e} + p_{e} \cos{\theta_{e}})}
		\end{equation}
		\begin{equation}
			Q^{2}_{QE} = 2 E_{\nu}^{QE} \left(E_{e} - p_{e} \cos{\theta_{e}}\right) - m_{e}^{2}
			\label{eq:q2}
		\end{equation}
		The cross sections are calculated in bins $i$ according to the following rule for example variable $\xi$, with $\epsilon$ representing signal acceptance, $\Phi$ the flux integrated over the energy range of the measurement, $T_{n}$ the number of targets (nucleons) in the fiducial region, $\Delta_{i}$ the width of bin $i$, and $U_{ij}$ a matrix correcting for detector smearing in the variable of interest:
		\begin{equation}
			\label{eq:dsigma}
			\left( \frac{d\sigma}{d\xi} \right)_{i} = \frac{1}{\epsilon_{i} \Phi T_{n} \left(\Delta_{i}\right)} \times \sum_{j}{U_{ij} \left(N_{j}^{\mathrm{data}} - N_{j}^{\mathrm{bknd\ pred}}\right)}
		\end{equation}
		
		We perform unfolding in these variables using a Bayesian technique\cite{D'Agostini unf} with a single iteration.  The unfolding matrices $U_{ij}$ needed as input are predicted by our simulation.  Our prediction for the neutrino flux $\Phi$ by which we then divide is derived from a GEANT4-based simulation of the NuMI beamline (described further in ref. \cite{antinumu PRL}).  In addition, the neutrino-electron elastic scattering measurement mentioned above provides an \textit{in situ}, data-based constraint for the flux estimate.  
		
\begin{figure}
			\centering
			\begin{adjustwidth}{-0.5in}{-0.5in}
				\subfigure[$\theta_{e}$]{%
					\includegraphics[width=0.41\textwidth]{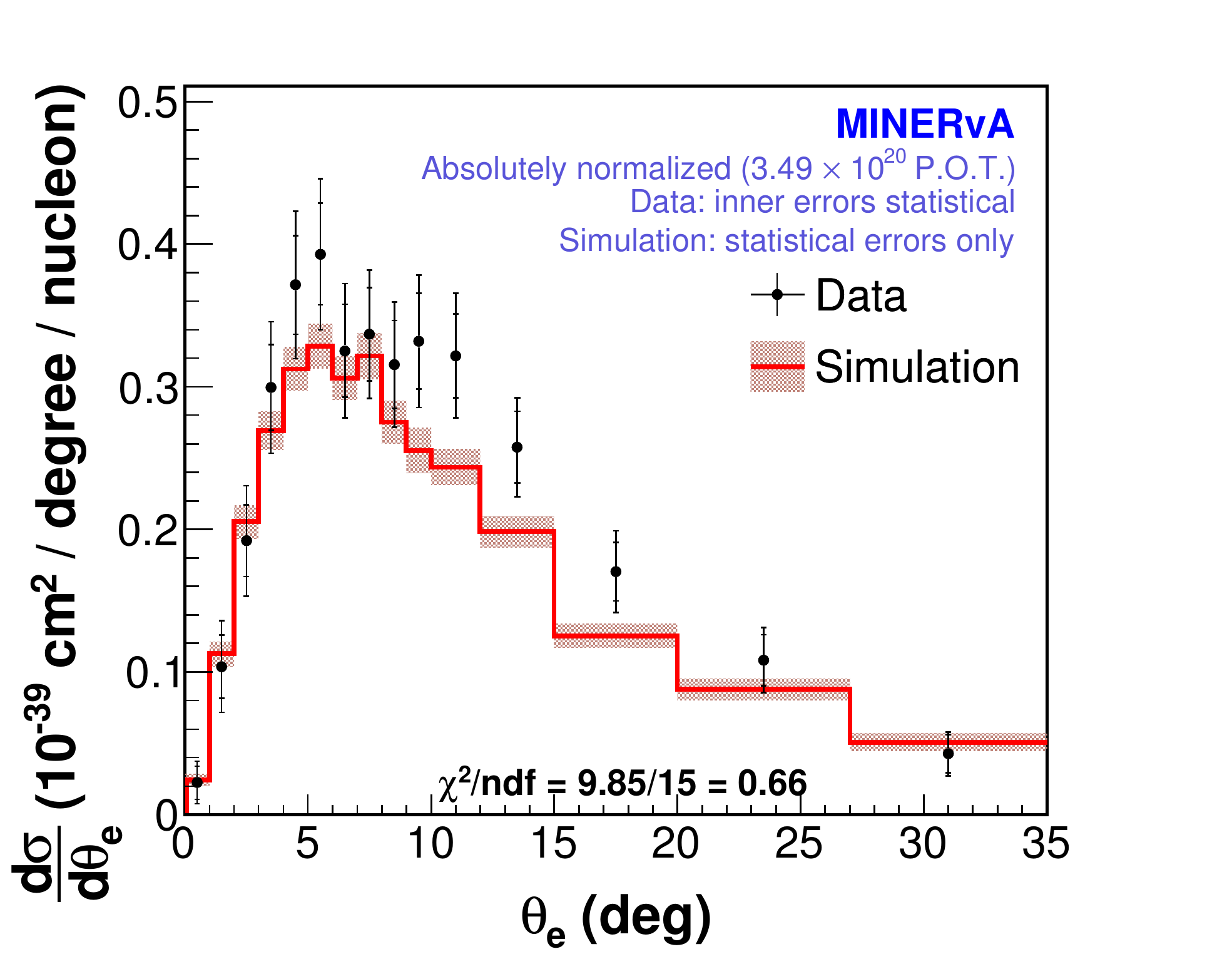}
				}%
				\subfigure[$E_{e}$]{%
					\includegraphics[width=0.41\textwidth]{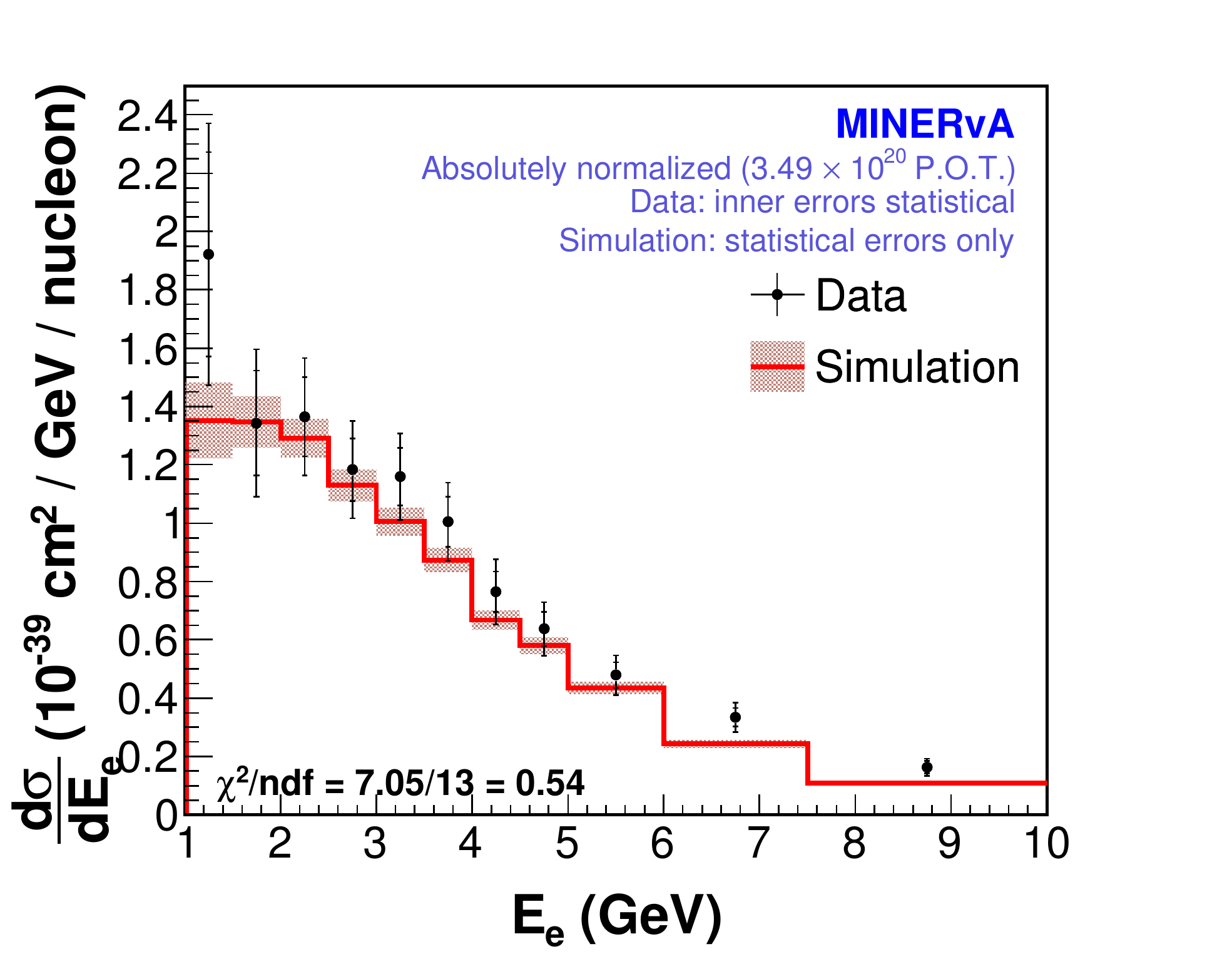}
				}%
				\subfigure[$Q^{2}_{QE}$]{%
					\includegraphics[width=0.41\textwidth]{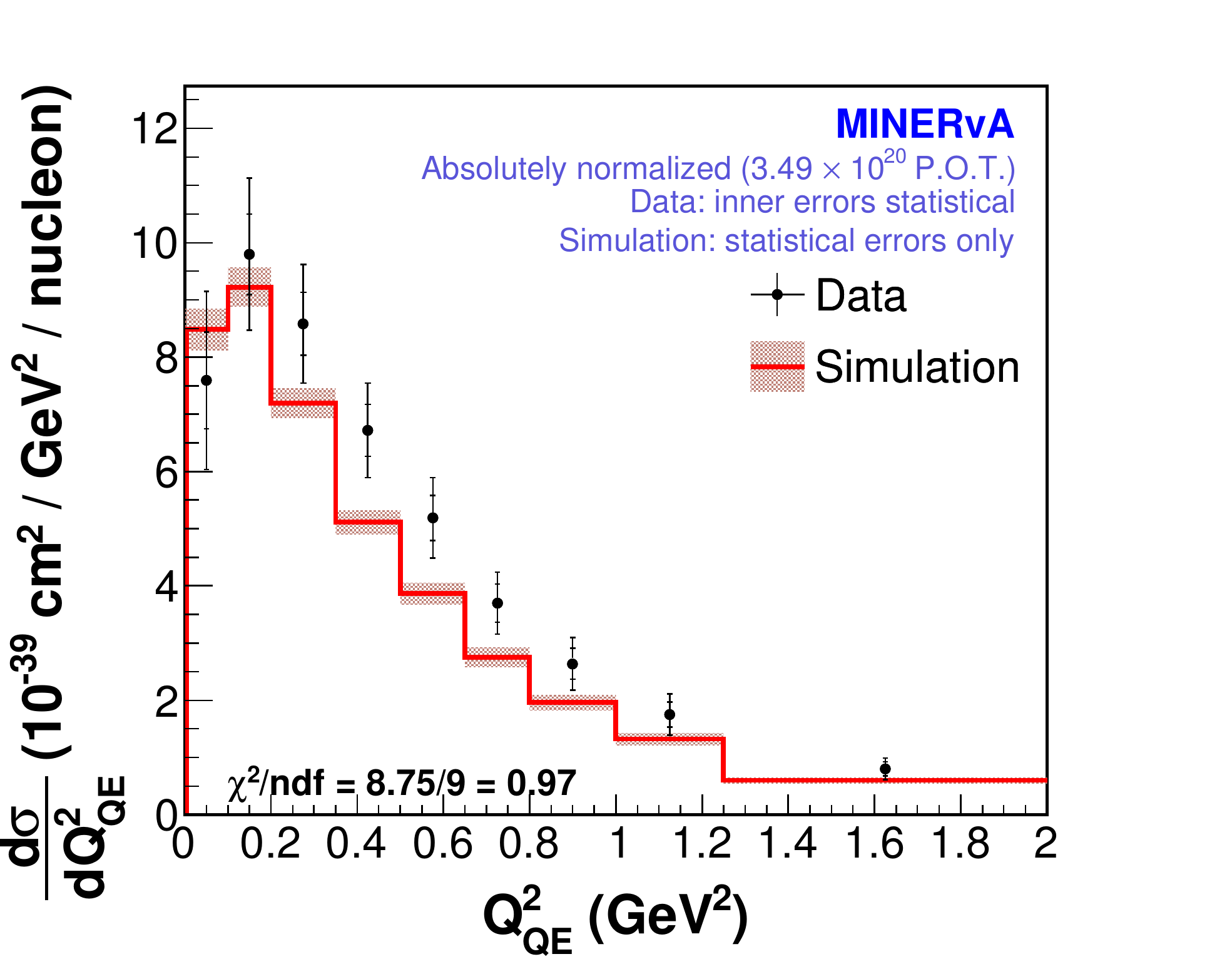}
					\label{fig:XS q2}
				}
			\end{adjustwidth}
			\caption{Differential cross sections.  Inner errors are statistical; outer are statistical added in quadrature with systematic.}
			\label{fig:XSs}
		\end{figure}

\begin{figure}
			\centering
			\includegraphics[width=0.75\textwidth]{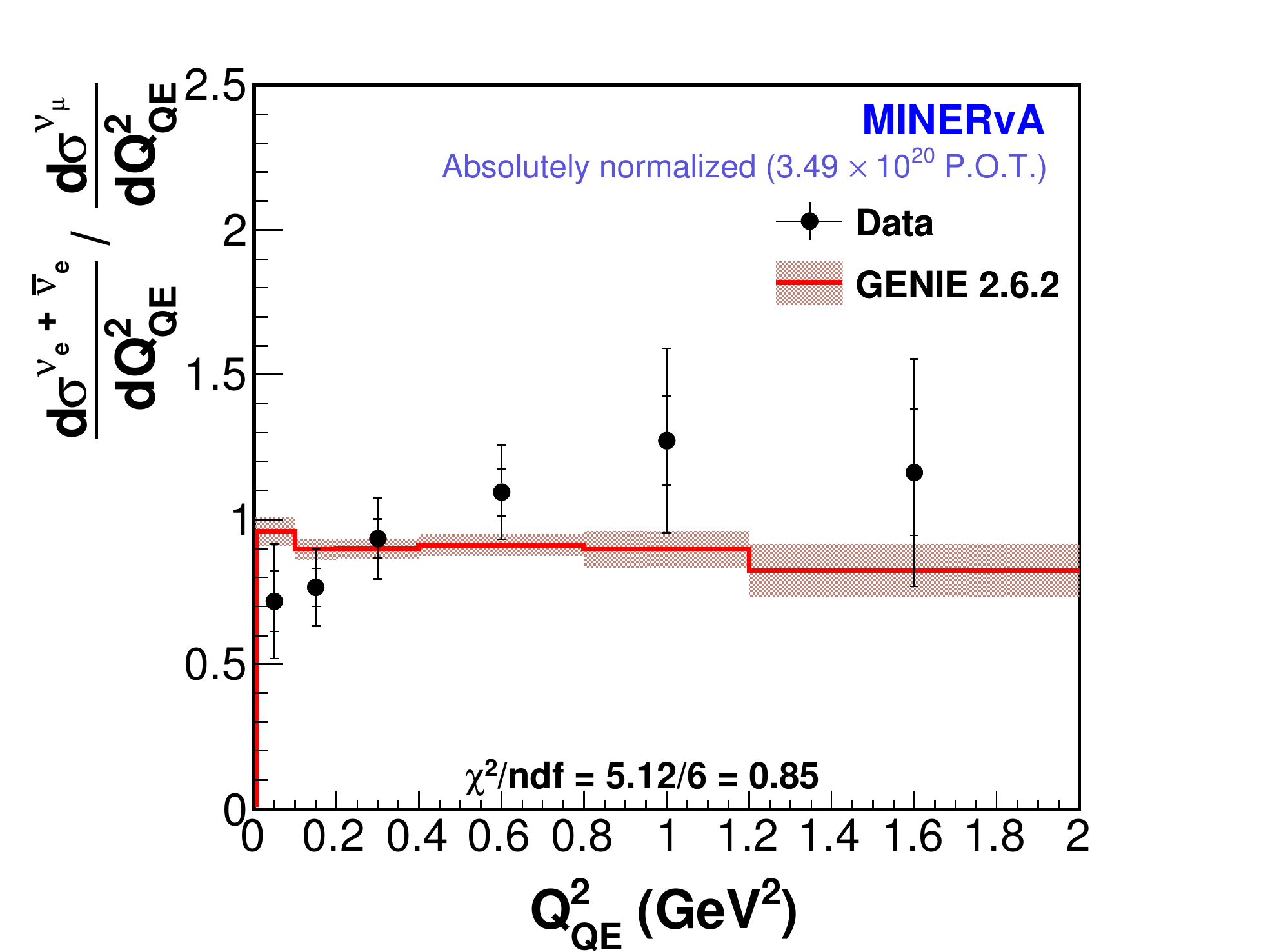}
			\caption{Ratio of $\frac{d\sigma}{dQ^{2}_{QE}}$ for $\nu_{e}$ to that for $\nu_{\mu}$.  Inner errors are statistical; outer are statistical added in quadrature with systematic.}
			\label{fig:ratio}
		\end{figure}

		The cross sections obtained from this procedure are given in fig. \ref{fig:XSs}.  To help understand whether any differences between the model and our data stem from deficiencies in the underlying cross section model itself (which is tuned to $\nu_{\mu}$ scattering data, as noted in the introduction) or differences between $\nu_{e}$ and $\nu_{\mu}$ interactions, we also computed the ratio of the cross section in fig. \ref{fig:XS q2} to a recent \mnv{} measurement of the same cross section for muon neutrinos, which is shown in fig. \ref{fig:ratio}.  We note that $Q^2$-dependent correlated errors we considered as part of the error treatment, such as that in the electromagnetic energy scale, can cause trends in the data similar to the difference between the prediction and observed shape in $Q^{2}$ in fig. \ref{fig:XS q2} and the apparent upward slope in fig. \ref{fig:ratio}.  When these correlated errors are taken into account, both the aboslutely normalized version shown here and the shape of the data distribution are consistent with the GENIE predictions within $1\sigma$; this is reflected in the $\chi^{2}$ values reported in the figures.
		
	\section{Conclusions}
		Though $\nu_{e}$ cross section data is vitally important for neutrino oscillation searches, experimental challenges have prevented extensive measurement of this quantity until recently.  In this first-ever measurement of $\nu_{e}$ CCQE scattering, we find that the electron neutrino cross section predictions of the GENIE generator, based on cross section models tuned to muon neutrino scattering data, are consistent with our measured values within our uncertainties.  This implies that the generator models in their current form are suitable for use by current neutrino oscillation experiments.  However, future experiments, which depend on significantly reducing the influence of cross section systematic uncertainties on their results, may require further data to resolve whether the apparent (but not significant) trends in our result correspond to real discrepancies between the models and nature.
		
		Furthermore, we observe a rare process not predicted by GENIE, neutral-current diffractive scattering from hydrogen, in a region adjoining the phase space occupied by electron neutrino events.  Detectors used in electron neutrino oscillation experiments that cannot differentiate between the two---such as water \v{C}erenkov devices---may require improvements to the available models in order to be sure their backgrounds are accurately predicted.


\begin{thebibliography}{99}
		\bibitem{T2K NIM} 
		K.~Abe {\it et al.} [T2K Collaboration],
		Nucl.\ Instrum.\ Meth.\ A {\bf 659}, 106 (2011)
		[arXiv:1106.1238].
  
		\bibitem{NOvA TDR} 
		D.~S.~Ayres {\it et al.} [NOvA Collaboration],
		FERMILAB-DESIGN-2007-01.
  
		\bibitem{Gargamelle nue}
		J.~Blietschau {\it et al.} [Gargamelle Collaboration],
		Nucl.\ Phys.\ B {\bf 133}, 205 (1978).
  
		\bibitem{T2K nue}
		K.~Abe {\it et al.} [T2K Collaboration],
		Phys.\ Rev.\ Lett.\  {\bf 113}, 241803 (2014)
		[arXiv:1407.7389].
		
		\bibitem{DayMcF}
		M.~Day and K.~S.~McFarland,
		Phys.\ Rev.\ D {\bf 86}, 053003 (2012)
		[arXiv:1206.6745].

		\bibitem{MINERvA NIM}
		L.~Aliaga {\it et al.} [MINERvA Collaboration],
		Nucl.\ Instrum.\ Meth.\ A {\bf 743}, 130 (2014)
		[arXiv:1305.5199].

		\bibitem{GiBUU FSI}
		O.~Lalakulich, U.~Mosel and K.~Gallmeister,
		Phys.\ Rev.\ C {\bf 86}, 054606 (2012)
		[arXiv:1208.3678].

		\bibitem{Martini corr}
		M.~Martini and M.~Ericson,
		Phys.\ Rev.\ C {\bf 87}, no. 6, 065501 (2013)
		[arXiv:1303.7199].
		
		\bibitem{Nieves corr}
		J.~Nieves, M.~Valverde and M.~J.~Vicente Vacas,
		Phys.\ Rev.\ C {\bf 73}, 025504 (2006)
		[hep-ph/0511204].

 		\bibitem{D'Agostini unf}
		G.~D'Agostini,
		Nucl.\ Instrum.\ Meth.\ A {\bf 362}, 487 (1995).
 		
		\bibitem{antinumu PRL}
		L.~Fields {\it et al.} [MINERvA Collaboration],
		Phys.\ Rev.\ Lett.\  {\bf 111}, no. 2, 022501 (2013)
		[arXiv:1305.2234].

		\bibitem{numu PRL}
		G.~A.~Fiorentini {\it et al.} [MINERvA Collaboration],
		Phys.\ Rev.\ Lett.\  {\bf 111}, 022502 (2013)
		[arXiv:1305.2243].
    		
		\bibitem{MINERvA nu+e} 
		J.~Park {\it et al.} [MINERvA Collaboration],
		arXiv:1512.07699.

  		\bibitem{MINERvA coherent}
		A.~Higuera {\it et al.} [MINERvA Collaboration],
		Phys.\ Rev.\ Lett.\  {\bf 113}, no. 26, 261802 (2014)
		[arXiv:1409.3835].
  
 		\bibitem{Lackner coh} K.~S.~Lackner, 
 		Nucl.\ Phys.\ B {\bf 153}, 526 (1979).
 		
		\bibitem{MB coh} A.~A.~Aguilar-Arevalo {\it et al.} [MiniBooNE Collaboration],
		Phys.\ Lett.\ B {\bf 664}, 41 (2008).
		[arXiv:0803.3423].
		
		\bibitem{Rein diffr}
		D.~Rein,
		Nucl.\ Phys.\ B {\bf 278}, 61 (1986).
	\end{thebibliography}
\end{document}